\documentclass{elsart}
\usepackage{epsfig}
\begin{document}

\begin{frontmatter}
  \title{Pulses in the Zero-Spacing Limit of the GOY Model}
  \author[isva]{K.H.  Andersen}, 
  \author[fi]{T. Bohr},
  \author[nbi]{M.H. Jensen}, 
  \author[nbi]{J.L. Nielsen} and
  \author[nbi]{P. Olesen} 
  \address[isva]{Department of Hydrodynamics
    and Water Resources (ISVA), the Danish Technical University,
    building 115, DK-2800 Lyngby, Denmark, tlf.: +45 45251420, 
    fax: +45 45932860, e-mail: ken@isva.dtu.dk, 
    www.isva.dtu.dk/$\tilde{\ }$ken.} 
  \address[fi]{Department of
    Physics, the Danish Technical University, building 309, DK-2800
    Lyngby, Denmark} 
  \address[nbi]{The Niels Bohr Institute,
    Blegdamsvej 17, DK-2100 Copenhagen {\O}, Denmark}

  \begin{abstract}
    We study the propagation of localised disturbances in a turbulent,
    but momentarily quiescent and unforced shell model (an
    approximation of the Navier-Stokes equations on a set of
    exponentially spaced momentum shells). These disturbances
    represent bursts of turbulence travelling down the inertial range,
    which is thought to be responsible for the intermittency observed
    in turbulence. Starting from the GOY shell model, we go to the
    limit where the distance between succeeding shells approaches zero
    (``the zero spacing limit'') and helicity conservation is
    retained. We obtain a discrete field theory which is numerically
    shown to have pulse solutions travelling with constant speed and
    with unchanged form.  We give numerical evidence that the model
    might even be exactly integrable, although the continuum limit
    seems to be singular and the pulses show an unusual super
    exponential decay to zero as $\exp(- \mathrm{const}\, \sigma^n)$
    when $n \rightarrow \infty$, where $\sigma$ is the {\em golden
      mean}. For finite momentum shell spacing, we argue that the
    pulses should accelerate, moving to infinity in a finite time.
    Finally we show that the maximal Lyapunov exponent of the GOY
    model approaches zero in this limit.
  \end{abstract}
\end{frontmatter}

Shell models of turbulence have increasingly been used as a laboratory
for testing hypotheses about the statistical nature of Navier-Stokes
turbulence. Even though they are derived by heuristic arguments they
are surprisingly adept at reproducing the statistical properties of
high Reynolds number turbulence. The reasons behind the apparent
success of the shell models still remains unclear.  In the present
paper an attempt will be made to examine the detailed behaviour of one
particular shell model -- the so-called GOY model by
Gledzer-Ohkitani-Yamada
\cite{Gledzer,OY,JPV,benzi,pisarenko,kada1,kada2,Lohse,bif,mogens} in
the limit where the ration between the wave number of successive shells
approach unity -- which we shall refer to as the zero-spacing limit.

One of the most important problems in turbulence is to understand {\em
intermittency} -- the very lumpy and irregular intensity of turbulent
motion.  Even in strongly turbulent states each region experiences
periods of almost complete quiescence interrupted by rapid burst
events, and this is believed to be the origin of the corrections to
Kolmogorov scaling seen experimentally, and predicted rather
accurately by the shell models \cite{JPV}.  Single burst motion
superimposed on the Kolmogorov spectrum (which emerges as a fixed
point of the shell model) have been studied in shell models by a
number of authors \cite{Siggia,pari:90,ditl:96,Dombre}.  In this
paper we also consider single burst motion but, in contrast to the
previous work, we do not superimpose a Kolmogorov background.  It was
shown recently, that prior to a burst in the GOY shell model, the
velocity field tends toward the ``trivial'' quiescent fixed point
\cite{Fridolin}. In this paper we thus study localised disturbances
propagating through the quiescent, unforced system.  One should keep
in mind in the following, that the shell models are defined in
$k$-space and that the pulses we describe are likewise in
$k$-space. Thus a pulse travelling to smaller $k$ should correspond
roughly to a spatial region of growing size, but we shall not go into
the detailed conversion of our results to real space, since this is
not even a completely well defined process in the context of (1-d)
shell models.

The paper is structured as follows: First a short introduction to the
GOY model is given in Section \ref{sec:GOY}. It was conjectured by
Parisi in a yet unpublished paper \cite{pari:90} that the correlation
functions of the GOY model (also called the structure functions) might
be calculated from the existence of soliton-like solutions on top of
the Kolmogorov fixed point $u(k) \propto k^{-1/3}$. For the study of the
pulses he proposed a continuum limit of the GOY model (the ``Parisi
equation''). Around the quiescent fixed point this equation does not
form pulses, but rather forms shocks. The Parisi equation around the
quiescent fixed point is studied in \cite{ande:98} and the results are
briefly reviewed in section \ref{sec:parisi}. Pulses in the quiescent
state are, however, seen in the zero-spacing limit of the GOY model.
The equation governing the zero-spacing limit is derived in section
\ref{sec:zero} and the qualitative behaviour of our model is
illustrated through numerical simulations.  In section
\ref{sec:pulses} a continuum version of the zero-spacing limit is
proposed, and a solution for a pulse is derived. The asymptotic
behaviour of a pulse is is treated in the case where the background
field is zero (section \ref{sec:asymp}) and non-zero (section
\ref{sec:const}). In section \ref{sec:acc} the acceleration of a pulse
which is seen away from the zero-spacing limit is examined. Finally
the Lyapunov exponent of the GOY model is calculated as a function of
the shell spacing, and it is shown that the model is not chaotic in
the limit of the shell spacing going towards zero (Section
\ref{sec:lyapunov})
\section{The GOY model}
\label{sec:GOY}
Shell models are formed by a truncation approximation (in $k$-space)
of the Navier-Stokes equations \cite{mogens}. The most well-studied
model is the ``GOY'' model of Gledzer-Ohkitani-Yamada.
 
This model yields corrections to the Kolmogorov theory \cite{JPV} in
good agreement with experiments \cite{Anselmet,vdWater}.  For the
``GOY'' shell model, wave-number space is divided into $N$ separated
shells each characterised by a wave-number $k_n=r^n \, k_0$ with
$n=1,\cdots ,N$.  Each shell is assigned a complex amplitude $u_n$
describing the typical velocity difference over a scale $\ell_n =
1/k_n$.  By assuming interactions only among nearest and next nearest
neighbour shells and phase space volume conservation one arrives at
the following evolution equations \cite{OY}
\begin{eqnarray}
  \lefteqn{(\frac{d}{ dt}+\nu k_n^2 ) \ u_n \ = }\nonumber \\
  & &  i \,k_n (a_n \,   u^*_{n+1} u^*_{n+2} \, + \, \frac{b_n}{r} 
  u^*_{n-1} u^*_{n+1} \, + \,
  \frac{c_n }{r^2} \,   u^*_{n-1} u^*_{n-2})  \ + \ f,
 \label{un}
\end{eqnarray}
with boundary conditions $b_1=b_N=c_1=c_2=a_{N-1}=a_N=0$.
$f$  is an external, constant forcing, and $\nu$ is the viscosity. 

An important property of the GOY model is that it is possible
to incorporate the conservations laws of turbulence as found
in the Navier-Stokes equations. The first conservation law that needs
to be satisfied is the conservation of energy. The energy in
the GOY model is given by
\begin{equation}
E ~=~ \sum_n |u_n|^2
\end{equation}
which should be conserved in the limits of no forcing and vanishing
viscosity.  This leads to the following relation between the
coefficients
\begin{equation}
a_n+b_{n+1}+c_{n+2}=0
\end{equation}
The constraint still leaves a free parameter $\delta$ so that one can
set $ a_n=1,\ b_{n+1}=-\delta,\ c_{n+2}=-(1-\delta)$ \cite{bif} (the
value of $a_n$ is fixed by the time scale).  Kadanoff {\em et al.}
\cite{kada1} observed that also a ``helicity'' invariant exists on the
form
\begin{equation}
H ~=~ \sum_n (-1)^n k_n |u_n|^2
\end{equation}
Conservation of this quantity leads to an additional constraint on the
parameter $b_n$
\begin{equation}
  b_n = \frac{1}{r} -1\quad \mathrm{or}\quad \delta = 1-\frac{1}{r}.
  \label{b}
  \label{equ:constraint}
\end{equation}
In the most studied case where $r=2$ one obtains the canonical values
$b_n = c_n = \frac{1}{2}$. In this paper we are going to focus on the
limiting case $r \to 1$ and according to Eq. (\ref{b}) $b_n
\rightarrow 0$ in order to satisfy helicity conservation.  The set
(\ref{un}) of $N$ coupled ordinary differential equations can be
numerically integrated by standard techniques.
\section{Some results on the Parisi continuum limit}
\label{sec:parisi}
In a previous paper \cite{ande:98} we studied a number of features of the
Parisi \cite{pari:90} continuum model. In particular, a ``burst'' event
(the evolution of an initial condition having support only in a finite
interval of $k$-space) was studied. The initial burst splits up into a
right and a left moving part and on the front of each part a shock is
formed. It was shown that the continuum equation without forcing or
dissipation can be explicitly written in characteristic form and that
the right and left moving parts can be solved exactly. When this is
supplemented by the appropriate shock condition it is possible to find
the asymptotic form of the burst.

In this section we shall give a brief review of some of the main
features previously obtained. The continuum Parisi equation is
obtained from the GOY model by letting the shell spacing approach
zero, and expanding to the first order.  By writing the distance
between the shells as $r=1+\epsilon$ with $\epsilon\ll 1$, we have
$k_n\approx \exp (n\epsilon)$, so with $n\sim$ const.$/\epsilon$ a
continuous range of values is obtained for the variable $k$.

To proceed we use a Taylor expansion of the type
\begin{eqnarray}
u_{n+1}(t)&=&u(k_{n+1},t)=u(\ln k_{n+1},t)\approx u(n
\ln(1+\epsilon),t)+
\ln (1+\epsilon)\frac{\partial u_n}{\partial \ln k}\nonumber \\
&\approx& u(k,t)+\epsilon k\frac{\partial u(k,t)}{\partial
k}\nonumber \\
&=&u(k,t)+\epsilon k\frac{\partial u(k,t)}{\partial k},
~{\rm with}~u(k,t)\equiv u(k_n,t)=u_n(t),
\end{eqnarray}
and similarly for $u_{n+2}, u_{n-1},$ and $u_{n-2}$.  To first order
one obtains \cite{pari:90}:
\begin{equation}
 u_t^*+\nu k^2 u^*=
  -ik(2-\delta)\left(u^2+3ku u_k \right),
\label{firstorder}
\end{equation}
where $u_k \equiv \partial u /\partial k$ etc..  By rescaling time with
$2-\delta$ we get the Parisi equation:
\begin{equation} u_t^* + 3 i k u u_k = - i k u^2,
\label{Parisi}
\label{equ:parisi}
\end{equation} 
for the case of no forcing or viscosity. Here subscripts denote
differentiation. Splitting into real and imaginary parts, $u=a+ib$, we
get
\begin{equation}
\label{a-Parisi} a_t -3ka b_k - 3k a b_k = 2 k a b
\end{equation} and
\begin{equation}
\label{b-Parisi} b_t - 3ka a_k + 3k b b_k = k (a^2 - b^2).
\end{equation} In numerical simulations we observed that a short while
after the initialisation, shocks are formed in both directions. As was
seen from the equations, the real part is only able to move to the
left, while the imaginary part moves both ways. Setting $a = 0$ in the
original equations, we get just a single real equation for the
inertial range:
\begin{equation} b_t + 3 k^2 b b_k= - k b^2
\label{equ:rpulse}
\end{equation} For the left-moving part of the pulse, the real and the
imaginary parts are seen to become proportional.  Inserting $b = Ca$
into the Parisi equation gives us two possibilities: $a=0$ (which we
already treated) and $a = \pm \sqrt{3} b$. For $a = \sqrt{3} b$ the
equation for the imaginary part becomes:
\begin{equation} b_t - 6 k^2 b b_k = 2 k b^2.
  \label{equ:lpulse}
\end{equation}

It is possible to explain the observed splitting of the pulse by
writing (\ref{Parisi}) in characteristic form. We shall not give the
details, since they can be found in the previous paper \cite{ande:98}.
The starting point is to rewrite (\ref{Parisi}) in a form resembling
the the Burgers equation
\begin{equation} v_t^* - i v v_x = 0,
\label{v-Parisi}
\end{equation} where $u = k^{-1/3} v$ and $k = (2x)^{-3/2}$. Of
course, there are three important differences to the Burgers equation:
Here we have an equation in $k-$space, $v$ is complex and the
conserved quantity is the energy $E=\int (2x)^{5/2}v^2$. Writing
$v=re^{i\theta/3}$ the Riemann invariants $J_{\pm}$ can be found, and
the solution can be expressed in terms of these
\begin{equation}
\label{r(J)} r= \left({\frac{J_{+}^3 + J_{-}^3}{2}}\right)^{1/3},
\end{equation} and
\begin{equation}
\label{theta(J)} \sin\theta = {\frac{J_{+}^3 - J_{-}^3}{J_{+}^3 +
J_{-}^3}}.
\end{equation} The formulation in terms of characteristics can now be
used to understand the splitting of the pulse found in Section
3.1. The Riemann invariants $J_\pm$ have the property that they are
constant along the curve:
\begin{equation}
  \frac{d x}{d t} = \pm r(x,t).
\end{equation}
The important point is that, since $r$ is nonnegative, one family
of characteristics ($J_+$) can never move to the left while the other
one ($J_-$) can never move to the right. If the initial condition has
support in a limited region of $x$, say $[x_-,x_+]$ the same is true
of $J_\pm$. They both vanish (in the initial condition) outside of
this interval. For times $t>0$ we compute the field values by finding
where, on the $x$-axis (i.e., $t=0$), the characteristics going through
the point $(x,t)$ emanate. Now, if $x>x_+$, the $J_-$-characteristic
going through this point must emanate from some $x_0 > x_+$ and thus
$J_-(x,t)= J_-(x_0,0) =0$. This means that either $r=0$ (which makes
the entire field vanish) or $\sin \theta = 1$ which means that
$\theta$ has the constant value $\pi/2$, which implies that $a=0$, in
agreement with the numerical simulations.

The splitting of the pulse into a right-moving part with $a=0$ and a
left-moving part with $a=\sqrt{3} b$ makes it possible to give a
complete solution for a single burst event. For the details, we refer
to \cite{ande:98}. Each pulse develops a shock which can be followed
by imposing conservation of energy across the shock.

The asymptotical solutions for the two pulses are:
\begin{eqnarray} 
  u_{right} &=& (2t)^{-2/3}k^{-1/3} \\ u_{left} &=&
  (tk)^{-1} \quad \mathrm{down\ to}\quad  k\sim C/t
\end{eqnarray} 
both decaying in time. Thus a burst created on top of the zero
solution not only travels down the inertial range until it is
dissipated by viscosity (here, by the continuum), it also travels
upwards, to the smallest $k$. The burst does not remain localised in
$k$-space, but is distributed over the $k$-range, and then decays. It
is interesting to note that the initial disturbance creates both a
forward and an inverse cascade. 

\begin{figure}[htbp]
%from ntplot.m
\begin{center}
  \epsfig{file=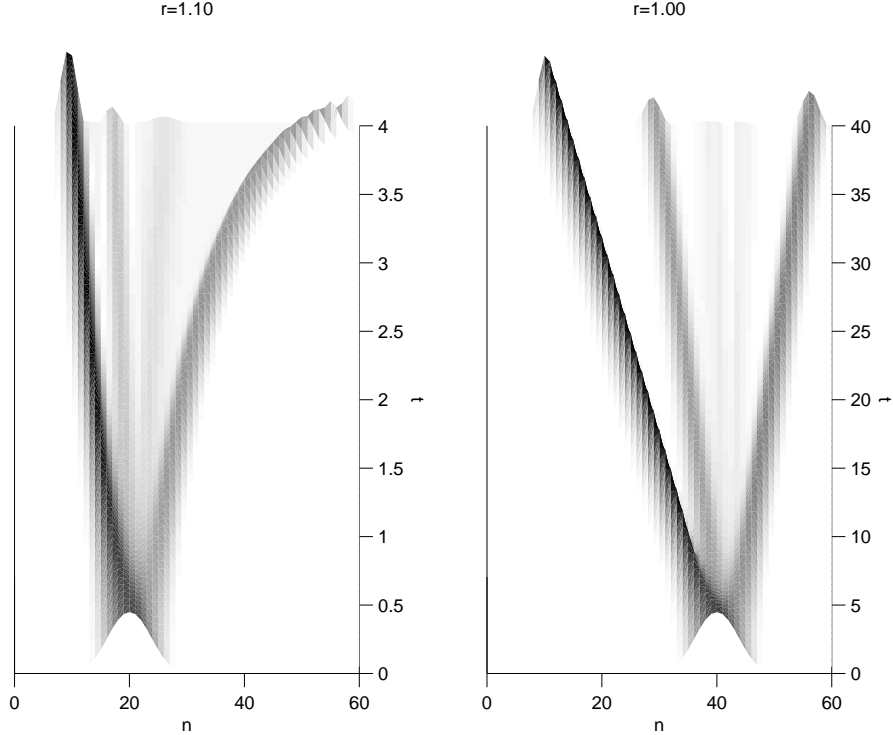, width=12cm}
  \caption{The development of a initial small bump in the GOY model
    with $r=1.1$ and $r=1.0$. Shown is $|u_n|$.}
  \label{fig:compare}
\end{center}
\end{figure}

\section{The zero-spacing limit}
\label{sec:zero}
When we study the GOY model for $r \ge 1$ we observe important
deviations from the behaviour of the Parisi equation described in the
previous section.  We focus on the motion on a single pulse in the
case where both the forcing and the viscosity is zero, i.e.,  $\nu = f
= 0$ and the two cases $r=1.0$ and $r=1.1$ (Fig. \ref{fig:compare}).
the initial condition is a localised bump:
\begin{equation}
  u_n = u_0 (1+i)\left\{ \cos [(n-n_0)\kappa_0 ] + 1 \right\}
\quad \mathrm{for} \quad x \in [n_0 \kappa_0 - \pi, n_0 \kappa_0 +
\pi],
  \label{ini}
\end{equation}
where $n_0$ is the center of the bump, $2\pi \kappa_0$ is the length
of the bump and $\frac{1}{2} u_0$ is the height of the bump.

The initial bump splits up into a left and a right moving
part with the same phases as in the Parisi equation. However, a shock
is not formed, rather the initial bump splits into a number of
individual pulses.  In the case $r=1.10$, the pulse to the right
accelerates whereas the pulse moving to the left decelerates, similar
to the shocks in the Parisi equation. In the case $r=1$, this does not
happen and both {\em the right- and left-moving pulses propagate
  with a constant velocity}. In both cases, however, the shapes
{\em stays invariant as time progresses}, completely similar to the
behaviour of a soliton in an integrable system, like the KdV equation.

%Let us reflect on the above observation is relation
%to intermittency. A standard way to observe intermittency is by
%measuring a velocity signal over a small length scale, the Kolmogorov
%length. If we imagine a disturbance starting at the large scales
%(small k-values), then it progates to larger k-values, finally
%reaching the Kolmogorov length. What we observe from Fig. is that the
%whereas the shape do not change much during this transport, it
%accelerates violently.  This means that the time for a burst to pass
%over k-values of the Kolmogorov length scale is extremely short,
%leading to a spiky, intermittent signal. In the limiting case ($r=1$),
%this does not occur, the final velocity is like the initial. The
%acceleration for $r>1$ is of course related to the well known decrease
%in time scale as the inertial range is passes.

%As mentioned earlier the limit $r=0$ is not completely regular. The
%above numerical solutions does however give some confidence that this
%limit can be explored with some profit. The two cases show the same
%phenomenology with an initial pulse splitting into left and right
%moving pulses. The pulses at $r=1.1$ and $r=1$ are quite similar,
%except for the dispersion relation.  The limit $r=1$ is rather special
%in a physical sence. In this limit the shell spacing is zero, such
%that even pulses travel in $n$-space they remain on the same point in
%$k$-space. As such, the limit is non-physical. This limit is however
%much easier to study than the $r>1$ as the pulses travel with constant
%velocity.

To be able to observe the pulses in more detail, we perform the limit $r
\rightarrow 1$ in a different way as is done when 
the Parisi equation is derived. We shall show that this leads to an
interesting model containing pulses moving with constant velocity and
which might even be a new example of a completely integrable
discrete field theory.

The starting point is the GOY model
\begin{equation} 
  ({\frac{du_n}{dt}} + \nu k_n^2 u_n)^* = - i k_n (
  u_{n+1} u_{n+2} -{\frac{\delta}{r}}u_{n-1}
  u_{n+1}-{\frac{1-\delta}{r^2}}u_{n-1} u_{n-2})
\end{equation} 
where $k_n = k_0 r^n$. We now introduce the variable
$w_n = k_{n-1} u_n$, and in terms of this variable we get
\begin{equation} 
  ({\frac{dw_n}{dt}} + \nu k_n^2 w_n)^* = - i (r^{-2}
  w_{n+1} w_{n+2} -\delta w_{n-1} w_{n+1}- (1-\delta) r^2 w_{n-1}
  w_{n-2})
  \label{equ:GOYw}
\end{equation} 
Using this substitution we get rid of the explicit dependence on $k_n$
on the right hand side of Eq. (\ref{equ:GOYw}).

Now we take the limit $r \rightarrow 1$ and at the same
time we invoke {\em helicity conservation}, which means that we must
choose $\delta = 1-1/r \rightarrow 0$. Finally, we are interested in
very large Reynolds numbers, so we let $\nu \rightarrow 0$ and then we
arrive at the model
\begin{equation} {\frac{dw^*_n}{dt}} = - i (w_{n+1} w_{n+2} - w_{n-1}
w_{n-2})
    \label{w*}
\end{equation} The middle term in the GOY model has disappeared and we
are left with the two terms directly responsible for the propagation of
the pulses to the left and the right.
\begin{figure}[tbp]
  % from ntplot2.m 
  \begin{center} 
    \epsfig{file=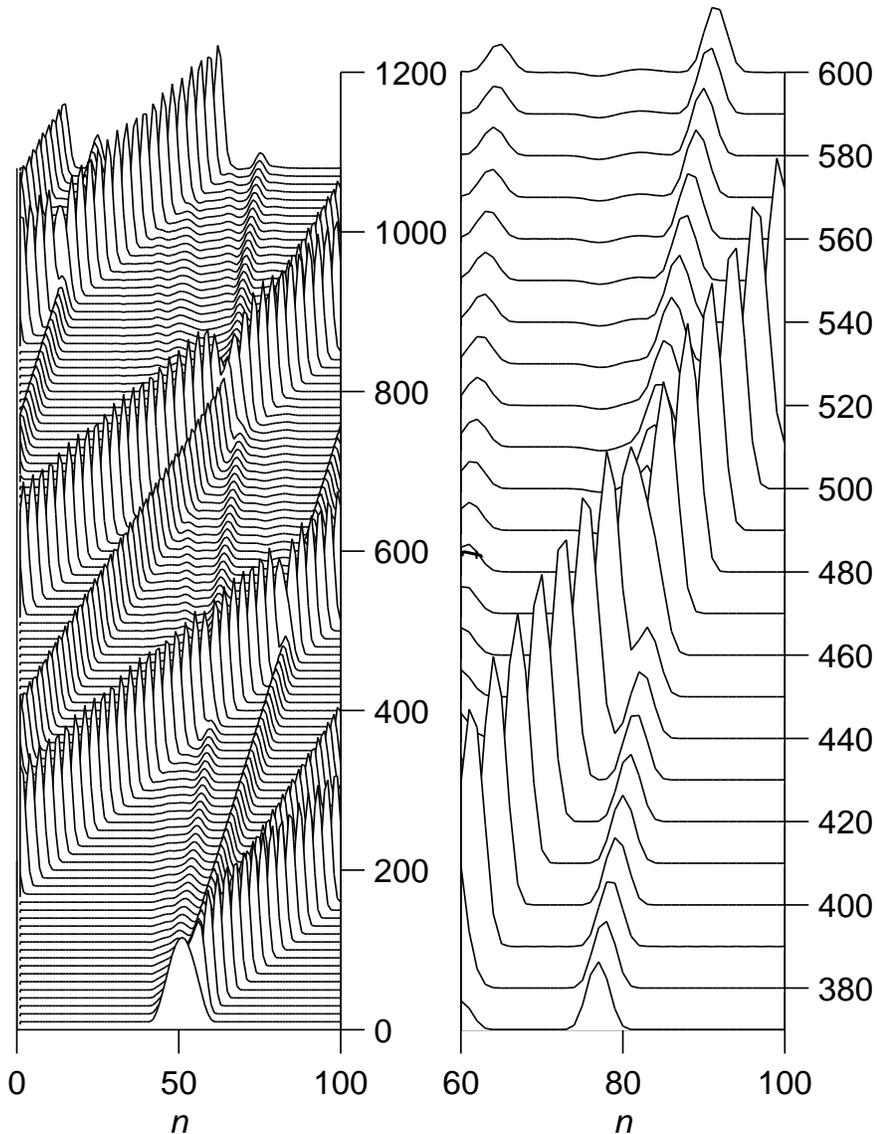,width=12cm} 
    \caption{Time-space plots of the imaginary part of the
      transformed GOY model Eq. (\protect\ref{equ:w}), with a
      cosine-bump as initial condition (\protect\ref{ini}). The
      boundary conditions are made periodic, to make it possible to
      see the interaction of the pulses.  The right picture is a zoom
      on an interaction between two pulses, where the time-delay in
      the collision process is clearly visible}
    \label{fig:ntplot2} 
  \end{center}
\end{figure}

This model has turned out to have very interesting properties. An
initial bump on zero background ($w=0$) splits into a series of left
and right moving pulses with constant velocity as already shown in
Fig. \ref{fig:compare}. Similar to the case for the Parisi equation
(\ref{equ:parisi}) the phases of the pulses only have certain fixed
values.  Let
\begin{equation} w_n = R_n e^{i \phi}
\end{equation} then
\begin{equation}
\label{R} {\frac{dR_n}{dt}} = (R_{n+1} R_{n+2} - R_{n-1} R_{n-2}) e^{i
(3 \phi - \pi / 2)}
\end{equation} and to have a solution with constant phase, the
exponential factor has to be real.  This means that $\sin (3 \phi -
\pi/2) = 0$ or
\begin{equation} 
  \phi = {\frac{2 p + 1}{6}} \pi;
\end{equation}
precisely the same values as for the Parisi equation \cite{ande:98}. For
these values of $\phi$, the exponential factor in (\ref{R}) becomes
$\cos (3 \phi - \pi/2) = \cos (p \pi) =(-1)^n$, giving rise to the
left (+1) and right (-1) moving fields respectively.  This behaviour
is in fact an intrinsic property of the inviscid GOY model, and stem
from the quadratic terms and the complex conjugation. This property of
selecting specific phases might explain the period-three organisation
of the pulses in the forced GOY model as noted by Okkels \& Jensen
\cite{Fridolin}.

The simplest case, on which we shall concentrate below, is $p=1$; 
a completely imaginary field
moving to the right. We thus let
$w_n = i \tilde{w}_n$ and dropping the tildes we get the real
equation
\begin{equation} 
 \frac{dw_n}{dt} = - w_{n+1} w_{n+2} + w_{n-1} w_{n-2}.  
 \label{equ:w}
\end{equation} 
The behaviour of (\ref{equ:w}) using a cosine-bump as initial
condition (\ref{ini}) is shown in Fig. \ref{fig:ntplot2}. The initial
bump splits into a number of pulses with the same shape but different
heights. The velocity of each pulse was measured to be $v = 2h$ where
$h$ is the height of the pulse. Two pulses can pass through each other
after a time delay produced by the collision (Fig.  \ref{fig:ntplot2},
right). These facts seem to indicate that the model (\ref{equ:w}) is
exactly integrable -- although, as we shall see later, probably not in
terms of analytic functions.

\begin{figure}[tbp]
  % Fra fit.m 
  \begin{center} 
    \epsfig{file=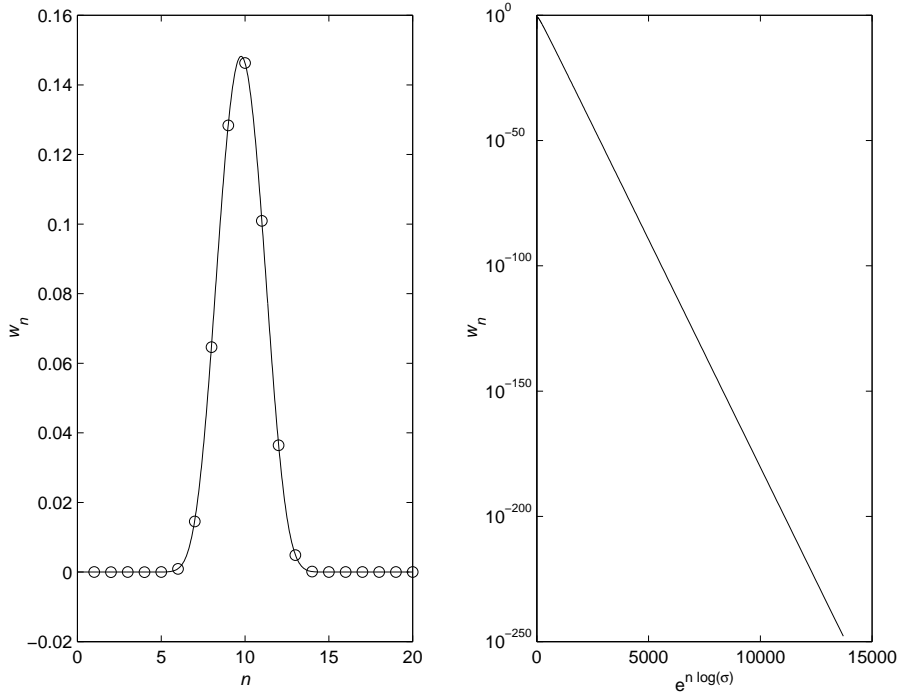, width=12cm}
    \caption{To the left is shown the shape of the pulse. The circles is
      the pulse extracted directly (at $t=210$ in Fig.
      \protect\ref{fig:ntplot2}) and the line is the pulse extracted
      using a time series from one point. On the figure to the right
      the horisontal axis is $z= \sigma^n$, where $\sigma = 1.618...$
      is the golden mean. Thus the tail of the pulse fits well to the
      form $w_n = e^{-a z}$.}
    \label{fig:pulse} 
  \end{center}
\end{figure}

One has to keep in mind that the system is discrete and therefore the
shape of a pulse deforms slightly in time. Despite this, one can
actually obtain a very precise shape function for the pulse, simply by
measuring the field $w$ continuously in time at a given point in the
lattice (Fig. \ref{fig:pulse})

The most characteristic feature of Eq. (\ref{w*}) compared to the
original GOY model, is the disappearance of the middle term. It should 
be noted that other variants of shell models display another limit. An 
example is the helicity-conserving shell models \cite{benz:95} where
in the $r=1$ limit the right term disappears, while the middle term is 
preserved. This makes the equation qualitatively different.

%A pulse moving to the right with constant velocity, say $v$, is a
%function only of the variable $x = n - v t$ and for such a function we
%get the simpler equation
%\begin{equation} 
%  -v {\frac{dw_n}{dx}} = w_{x+1} w_{x+2} - w_{x-1} w_{x-2}
%\end{equation} where $x$ is now a continuous variable, although the
%right hand side is still explicitly discrete. The velocity $v$ can be
%scaled away by letting $w \rightarrow v w$ and thus we finally arrive
%at the parameter-less pulse equation
%\begin{equation}
%  \label{wp} -\frac{dw_n}{dx} = w_{x+1} w_{x+2} - w_{x-1} w_{x-2}
%\end{equation}
%(which explains the linear dependence on the pulse height found above).

\section{Pulses in the continuum limit} 
\label{sec:pulses}
We shall now show that the
continuum limit of (\ref{equ:w}) is non-trivial and thus that the form of
the pulses might be unusual.  By expanding $w$ in a Taylor-series and
retaining, on the right-hand-side of (\ref{equ:w}) terms $w^{(p)}
w^{(q)}$ with $p+q \le 3$ we get the continuum limit
\begin{equation} 
\label{exp3}
  -w_t= 6 w w_{x} + 6 w_{x} w_{xx} + 3 w w_{xxx}
\end{equation} where again the subscripts denote differentiation.

Using the identity $(w^2)_{xxx} = 6 w_{x} w_{xx} + 2 w w_{xxx} $ the
continuum model can be written:
\begin{equation}
\label{cont1} -w_t= 6 w w_{x} + (w^2)_{xxx} - w w_{xxx}
\end{equation}
To study the asymptotics, we shall start by omitting the last
term. The resulting equation
\begin{equation}
\label{QKdV} -w_t = 6 w w_{x} + (w^2)_{xxx}
\end{equation} 
strongly resembles the Korteweg de Vries (KdV) equation. We shall
refer to it as the Quadratic Korteweg de Vries (QKdV) equation.

\subsection {Solitons in the KdV equation}
Let us briefly review how the solitons of the KdV equations are found.
We look for pulse solutions of the KdV equation
\begin{equation}
\label{KdV} w_t = w w_{x} + w_{xxx}
\end{equation} which are only a function of $x \rightarrow x + v t$,
so that (\ref{KdV}) becomes
\begin{equation}
\label{vKdV} v w' = w w' + w'''
\end{equation} 
Now we can integrate once to get
\begin{equation}
  \label{IKdV} w''-v w + {\frac{1}{2}} w^2 =C_1
\end{equation} 
Since we are interested in a pulse where the LHS approaches 0 as $x
\rightarrow \pm \infty$ we must take $C_1 =0$. The resulting equation
is that of a Newtonian particle moving in time $x$ in the potential
\begin{equation}
\label{pot} V(w) = \int_0^w (-v z + {\frac{1}{2}} z^2) dz = 
-{\frac{1}{2}}v w^2 +{\frac{1}{6}} w^3
\end{equation} Energy conservation has the form
\begin{equation}
\label{E} {\frac{1}{2}} (w')^2 + V(w) = C_2
\end{equation} Again the boundary conditions dictate that $C_2=0$ and
we can solve (\ref{E}) as
\begin{equation}
\label{sol} x(w)-x(w_0) = \int_{w_0}^w {\frac{dz}{\sqrt{-2 V(z)}}}
\end{equation} This leads to the $sech^2$ for  $w(x)$ but for our
purposes the most important point is that the soliton corresponds to
the homoclinic orbit starting from the potential hill-top $w=0$ at $x
\rightarrow -\infty$ and returning back at $x \rightarrow -\infty$. The
quadratic nature of the hill-top means that the integral for $x(w)$
{\em diverges} logarithmically for $w \rightarrow 0$ and thus the soliton
extends to infinity with an exponential tail in both directions.

\subsection{Solitons in the QKdV equation on a vanishing background}
We now look for a pulse solutions of (\ref{QKdV}) and let $x
\rightarrow x -v t$, so that (\ref{QKdV}) becomes
\begin{equation}
\label{vKKdV} v w' = 6 w w' + (w^2)'''
\end{equation} 
Again we can integrate once to get
\begin{equation}
\label{IQKdV} (w^2)'' - v w + 3 w^2 =C_1
\end{equation} 
and, if at some (finite or infinite) point the left hand side
vanishes, we must take $C_1=0$. This means that we are looking for
pulses with $w\rightarrow c = 0$. To make a Newtonian problem out of
this we now introduce the variable $y=w^2$. Then we get
\begin{equation}
\label{IQKdVy} y'' - v \sqrt{y} + 3 y = C_1
\end{equation} which is potential motion in the potential
\begin{equation}
\label{Kpot} V(y)= - {\frac{3}{2}}b y^{3/2} +{\frac{3}{2}} y^{2}
\end{equation} where $b=4 v/9$ and the total energy (\ref{E}) must
again be 0. Thus the solution can again be written as (\ref{sol}), but
this time the integral converges at $w=0$ so we can write explicitly
for the solution which vanishes at $x=0$:
\begin{equation}
\label{Ksol} x(w) = \int_{0}^y {\frac{ds}{\sqrt{-2 V(s)}}} = \frac{2}{\sqrt{3}}
\int_0^w {\frac{dz}{\sqrt{b z - z^2}}} = \frac{2}{\sqrt{3}} 
\cos^{-1}{\frac{b-2w }{b}}
\end{equation} which can be inverted as
\begin{equation}
\label{Ksol1} w = {\frac{b}{2}}(1-\cos \frac{\sqrt{3}x}{2}) = \frac{2 v}{9}(1-\cos \frac{\sqrt{3}x}{2})
\end{equation}

We verify that $w(0) = 0$ but now the solution has support only on the
interval $ [0,4 \pi]$. Outside of that it is identically zero which
means that there is a shock in $w_{xx}$ at those points. The strength
of this shock will presumably diminish when higher order
approximations to (\ref{equ:w}) is used. But one can easily see that
the term $w w_{xxx}$ which we omitted from (\ref{cont1}) will not
change this fact (although it will probably alter the simple solution
(\ref{Ksol1})).  Closely to the right of $x=0$, $w(x) \approx
\frac{v}{12} x^2$.  Thus $w_x$ and $(w^2)_{xxx}$ both go linearly in
$x$ (and balance), whereas $(w^2)_x$ and $w w_{xxx}$ both go like
$x^3$ and do not alter the asymptotics for $x\rightarrow 0^+$. This
will be verified explicitly in the next section.

\subsection{ Higher approximations.}
It is possible to take the term $w w_{xxx}$, which we omitted in
(\ref{exp3}) explicitly into account \cite{Ola:98}. Multiplying
(\ref{exp3}) with $2w$ on both sides, we can rewrite this as:
\begin{equation}
-(w^2)_t = 4 (w^3)_x + 6 (w^2w_{xx})_x
\end{equation}
If we again assume an uniform propagation with constant velocity $v$,
then $w(x,t)$ is only a function of $x-vt$, and the equation above
reduces to
\begin{equation}
v(w^2)' = 4 (w^3)' + 6(w^2w'')'
\end{equation}
which can immediately be integrated to yield:
\begin{equation}
w^2(-v+4w+6w'') = {\rm constant} 
\label{no-soliton}
\end{equation}
The non-trivial solution must therefore satisfy
\begin{equation}
6w''+4w = v
\end{equation}
with solution in the form of the ``compact'' soliton
\begin{equation}
w = {\frac {v}{4}}(1-\cos\sqrt{\frac{2}{3}}x)
\end{equation}
analogously to (\ref{Ksol1}) but no solutions decaying to
$0$ at infinity.

This approach is tied to the existence of a 
conserved energy given by: $E=\int dx\, w(x,t)^2$, which is conserved 
only if $2ww_t$ can be written as a total derivative in $x$:
\begin{equation}
2ww_t = F(w,w_x,..)_x
\end{equation}
But if we furthermore 
assume a dependence only on $x-vt$, then the
above equation reduces to an equation of total derivatives:
\begin{equation}
-v(w^2)' = F(w,w',..)'
\end{equation}
which can be integrated to give: 
\begin{equation}
F(w,w',..) + vw^2 = {\rm constant.}
\end{equation}
To have finite-energy solitons, $w$ must go to zero at infinity. If 
furthermore $F(w,w',..)$ goes to zero as $w$ goes to zero, 
the constant must be zero. We can then divide by $w^2$ and the
equation reduces to
\begin{equation}
\frac{F(w,w',..)}{w^2} = -v
\label{solution}
\end{equation}

Keeping terms up to $5$'th order in the derivatives in equation
(\ref{equ:w}) in the continuum limit, we get:
\begin{equation}
-w_t = 6 ww_x+3ww_{xxx}+6w_xw_{xx}+\frac{11}{20}ww_{xxxxx}
+\frac{3}{2}w_xw_{xxxx}+2w_{xx}w_{xxx}.
\end{equation}
Miraculously, this can be rewritten as
\begin{equation}
-(w^2)_t = \left( 4 (w^3) + 6 (w^2w_{xx}) + \frac{11}{20}w^2w_{xxxx}
+\frac{2}{5}ww_xw_{xxx}-\frac{2}{5}w_x^2w_{xx}+\frac{4}{5}ww_{xx}^2\right)_x
\end{equation}
so that 
\begin{equation}
F(w,w',..) = - 4w^3 - 6w^2w_{xx} - \frac{11}{20}w^2w_{xxxx}
- \frac{2}{5}ww_xw_{xxx} + \frac{2}{5}w_x^2w_{xx} - \frac{4}{5}ww_{xx}^2
\end{equation}
Equation (\ref{solution}) is then:
\begin{eqnarray}
v & = & 4w+6w''+\frac{11}{20}w''''
+\frac{2}{5}\frac{w'w'''+2w''^{\,2}}{w}-\frac{2}{5}\frac{w'^{\,2}w''}{w^2}
\nonumber\\
& = & 4w+6w''+\frac{11}{20}w''''+\frac{2}{5}\left(\frac{w'^{\,2}w''}{w}
\right)'\frac{1}{w'} 
\end{eqnarray}
which might have non-trivial soliton-solutions as long as the last
term equals $v$ at large $x-vt$. So for large $x-vt$, we must have:
\begin{equation}
\frac{5}{2}vw' = \left(\frac{w'^{\,2}w''}{w}\right)'
\end{equation}
or, by integration:
\begin{equation}
\frac{5}{2}vw = \frac{w'^{\,2}w''}{w} - C
\end{equation}
where C is some integration-constant. Now multiply this with $ww'$:
\begin{equation}
\frac{10}{3}v(w^3)'+ 2C(w^2)' = (w'^{\,4})'
\end{equation}
and integrate again
\begin{equation}
\frac{10}{3}vw^3+ 2C w^2 = w'^{\,4}
\end{equation}
where no extra constant arises, because of the boundary-condition at
infinity. Thus
\begin{equation}
w' = \left(\frac{10v}{3}w^3 + 2Cw^2\right)^{1/4}  
\end{equation}
If $C\neq 0$, the 2nd term dominates asymptotically, and we get
$w(x,t) \sim (x-vt)^2$ for large $x-vt$. On the other hand if $C=0$
then $w(x,t) \sim (x-vt)^4$ for large $x-vt$. In both cases $w(x,t)$
does not decay to zero as $x-vt$ goes to infinity. So again there
are no soliton-solutions, that decay to zero at infinity. 

\subsection{ Solitons in the QKdV with $c >0$.}
The strange appearance of the solitons is closely linked to the fact
that $w$ decays to zero asymptotically. If $w\rightarrow c > 0$ for $x
\rightarrow \pm \infty$, the pulse will decay exponentially to this
value far away just as in the KdV case. The difference is that the
constant $C_1$ now no longer vanishes. Instead $C_1 = -v c + c^2$ and
the potential acquires the form
\begin{equation}
\label{Kpotc} V(y)= - {\frac{1}{2}}b y^{3/2} +{\frac{1}{2}} y^{2} -
C_1 y
\end{equation} Now, for $v>0$ and small $c>0$, $C_1 \approx -v c <0$ 
so the potential acquires a small hill top at $w=c$. This hill top is
non singular (albeit in a very small region) and we get back to the
usual situation (like KdV) where the soliton spreads over the entire
region with an exponential tail. Near $w=c$ we get from (\ref{sol})
\begin{equation}
\label{Ksolc} x(y)-x(y_0) = \int_{y_0}^y {\frac{ds}{\sqrt{- (s-c)^2
V''(c)}}} = {\frac{1}{\sqrt{- V''(c)}}}\log {\frac{y-c}{y_0-c}}
\end{equation} where $V''(c)=-{\frac{v}{2 c}} $. Thus the solution
decays as
\begin{equation}
\label{Ksolc1} y(x) = w^2(x) =c+ (y_0-c) e^{\alpha x}
\end{equation} where $\alpha = \sqrt{- V''(c)} \sim
(\sqrt{c})^{-1}$. Thus the decay length $\xi \sim \sqrt{c}$.

\section{Asymptotics of solitons in the discrete model}
\label{sec:asymp}
The results obtained in the continuum limit in the previous section
are not born out by simulations. The pulses do not seem to have
compact support, but decay far away. The decay in not exponential, but
seems to more rapid.  We shall now give an estimate of this decay and
subsequently try to refine it.

Let us approximate the derivative in (\ref{equ:w}) as $w'(x) \approx
w(x+1)-w(x)$.  Further, expecting a rapid decay, we shall neglect
$w(x+1)$ and $w(x+2)$ compared to $w(x)$, $w(x-1)$ and $w(x-2)$. Thus
we simplify (\ref{equ:w}) to
\begin{equation} w(x) = w(x-1) w(x-2)
\end{equation}

Now let $\zeta(x) = \log(x)$ whereby we find
\begin{equation}
\label{Fib} \zeta (x) = \zeta (x-1) + \zeta (x-2)
\end{equation} This recursion relation is identical to the one for the
Fibonacci numbers, and in general $\zeta(x)$ will grow like
$\sigma^x$. To determine $\sigma$, assume that
\begin{equation} \zeta (x) = a \sigma^x
\end{equation} which should be valid at large $x$. Inserting into
(\ref{Fib}) gives us
\begin{equation} \sigma^2-\sigma -1=0
\end{equation} with the (positive) solution
\begin{equation}
\label{gm} \sigma = {\frac{1}{2}} (1+\sqrt{5})
\end{equation} the so-called {\em golden mean}. This gives the
extremely rapid decay
\begin{equation} w(x) \approx e^{-a \sigma^x}
\end{equation} with some positive constant $a$. This rather unusual
behaviour is extremely well represented by the numerical solution
(Fig. \ref{fig:pulse}).

To refine this, assume that
\begin{equation} w(x) = f(x) e^{a \sigma^x}
\end{equation} Inserting this into (\ref{equ:w}) and cancelling a factor
of $e^{a \sigma^x}$ (using (\ref{gm})) we get
\begin{equation} a f(x) \sigma^x \log \sigma - f'(x) = f(x-1) f(x-2) -
f(x+1) f(x+2) e^{-d a \sigma^x}
\end{equation} where $d =\sigma + \sigma^2 - \sigma^{-1} -
\sigma^x{-2} \approx 3.2 >0$. If the function $f(x)$ is ``reasonable''
the last term can be dropped and we get
\begin{equation}
\label{sd} a f(x) \sigma^x \log \sigma - f'(x) = f(x-1) f(x-2)
\end{equation} and we can get the asymptotic behaviour $f \sim
\sigma^x $ by matching the first term on the left hand side to the
right hand side. Thus $f(x) \approx c \sigma^x $ with $c= a \sigma^3
\log \sigma$ and the solution takes the form
\begin{equation} w(x) = a \sigma^{(x+3)} \log \sigma e^{a \sigma^x} =a
\log \sigma e^{(x+3) \log \sigma + a \sigma^x}
\end{equation} Note that $a$ is still not determined and we presumably
need to solve for the whole soliton structure to get this constant.

\begin{figure}[htbp]
  \begin{center}
    \epsfig{file=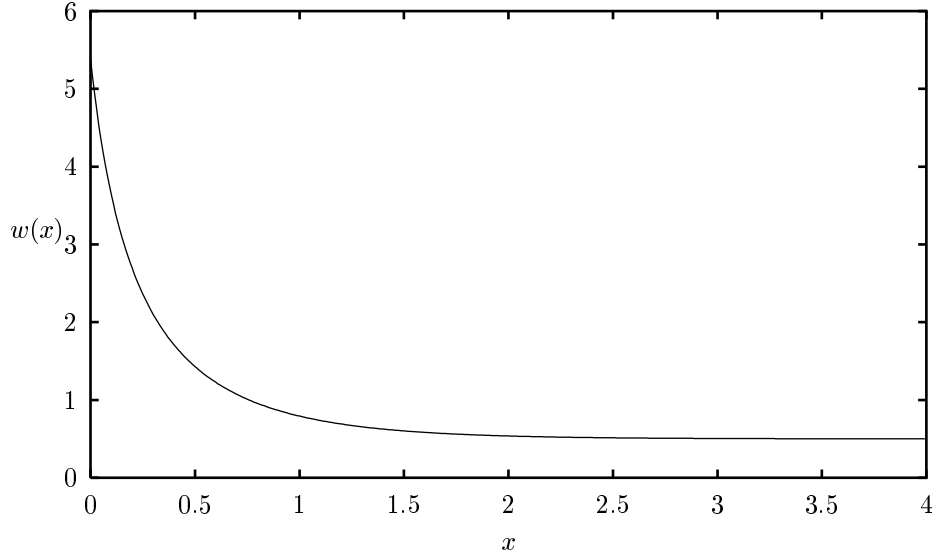} %\input{fig.pstex}
    \caption{An example of an ``exact'' solution of a pulse using
      the expansion
      Eq. (\protect\ref{ansatz}) with the coefficients obtained
      by the recursion relation Eq.
      (\protect\ref{recursion}) with $k=2$, $c_0=0.5$ and $c_1=2$.
      Note that $w_n$ asymptotically approaches $c_0$.}
    \label{fig:fig}
  \end{center}
\end{figure}

\section{Expansion for non-zero background}
\label{sec:const}
As seen in section \ref{sec:asymp} pulses converging to $c_0 > 0$
decay exponentially. This is born out by out numerics and we now give
a method to obtain the asymptotic behaviour of a pulse solution.

We consider
\begin{equation}
\dot{w}_n=w_{n+1}w_{n+2}-w_{n-1}w_{n-2}.
\end{equation}
For the field $w_n(t)$ we make the expansion
\begin{equation}
w_n(t)=\sum_{s=0}^{s=\infty}c_s\exp (-s(kn-vt)).
\label{ansatz}
\end{equation}
Hence we have
\begin{equation}
v\sum_1^\infty lc_l\exp (-l(kn-vt))=-2\sum_{l=1}^\infty \exp (-l(kn-vt))
\sum_{s=0}^l c_s c_{l-s}\sinh (2lk-sk).
\end{equation}
Thus we have the recursion relation
\begin{equation}
c_l(vl-2c_0\sinh (2lk)-2c_0\sinh (lk))=-2\sum_{s=1}^{l-1}c_sc_{l-s}
\sinh (2lk-sk).
\label{recursion}
\end{equation}
Although the recursion relation is non-linear, it turns out that it can
be solved successively in terms of the two constants $c_0$ and $c_1$.
For $l=1$ we get
\begin{equation}
v=2c_0(\sinh k+\sinh 2k),
\end{equation}
giving a relation between the constant background $c_0$, the width
$k$, and the velocity $v$. For $l=2$ we obtain
\begin{equation}
c_2=c_1^2\frac{2\sinh 3k}{2v-2c_0\sinh 2k-2c_0\sinh 4k}.
\end{equation}
Similarly, for $l=3$ we get
\begin{equation}
c_3=4c_1^3\frac{(\sinh 5k+\sinh 4k)\sinh 3k}{(3v-2c_0\sinh 6k-2c_0\sinh 3k)(2v-
2c_0\sinh 4k-2c_0\sinh 2k)}.
\end{equation}
It is straightforward, but tedious, to continue this calculation (we have 
computed $c_4$ and $c_5$).

To show that (\ref{ansatz}) is a solution, we need to show that the 
coefficients $c_l$ obtained from the recursion relation (\ref{recursion})
are such that the sum in (\ref{ansatz}) is either convergent or Borel
summable. We have not succeeded in this in general. However, it
is possible to show convergence when the width $k$ is large, and also
presumably when $k$ is small. Let us consider $k$ to be large. The velocity
then becomes
\begin{equation}
v\approx c_0e^{2k}+O(e^k).
\end{equation}
>From the recursion relation (\ref{recursion}) we get a solution of the form
\begin{equation}
c_l\approx c_1^l e^{-k(l-1)}/c_0^{l-1}(1+O(e^{-k})).
\end{equation}
The corresponding asymptotic solution can be found by performing the
sum in (\ref{ansatz}),
\begin{equation}
  w_n(t)\approx\frac{c_0+c_1 e^{-kn+c_0t\exp 2k}}{1-(c_1/c_0)e^{-k(n+1)
      +c_0t\exp 2k}}.
  \label{solution2}
\end{equation} 

Numerically, one can easily generate 
an ``exact'' solution by
Eq. (\ref{ansatz}) using the recursion relation Eq. (\ref{recursion}).
Of course the solution exists only if the coefficients $c_s$ converge.
We ohave observed that the radius of convergence in parameter space
has a ``hole'' around the point $c_0 = 0, c_1 = 0$. This is not surprising
because this expansion is only valid on non-zero background and from 
the structure of the recursion relations it is clear that $c_0 = c_1 =0$
leads to a trivial case where $c_s =0$ for all $s$. Away, from the open
set around $c_0 = c_1 =0$ we find a very fast convergence of the series.
An example of the obtained solution is shown in figure \ref{fig:fig}.

Note that formally we can apply this expansion also to the case
treated in section \ref{sec:const}, where $w \rightarrow 0$ for $r \rightarrow
\infty$: At first sight this seems to require $c_0=0$, which doesn't
work. However one has to keep in mind that the results of the previous
section shows that in this case $k$ must be negative, indeed $k=-log
\sigma$ and this means that we have an expansion in terms of {\em
  growing} exponentials. The background value $c$ is therefore not
given simply by $c_0$ in this limit, but all higher terms will
contribute to it.

\section{Acceleration of solitons for $r>1$}
\label{sec:acc}
When $r$ is slightly above unity, the main difference between our
zero-spacing model and the real GOY model seems to be the acceleration
of the bursts seen e.g. in Fig. \ref{fig:compare}. Letting $r= 1+
\epsilon$ in (\ref{equ:GOYw}) and retaining helicity conservation ($
\delta = 1-1/r \approx \epsilon$) we see that our inviscid model
(\ref{w*}), in lowest order of $\epsilon$, becomes
\begin{equation} {\frac{dw^*_n}{dt}} = - i (w_{n+1} w_{n+2} - w_{n-1}
w_{n-2} -4 \epsilon w^2_{n})
    \label{w*eps}
\end{equation}
For purely imaginary $w$ this is
\begin{equation} 
 \frac{dw_n}{dt} = - w_{n+1} w_{n+2} + w_{n-1} w_{n-2} + 4 \epsilon w^2_{n} 
 \label{equ:weps}
\end{equation}
In the continuum limit, neglecting all higher order derivatives we get
\begin{equation} 
 \frac{\partial w }{\partial t} = - 6 w w_x + 4 \epsilon w^2 
 \label{equ:conteps}
\end{equation}
from which we clearly see the acceleration caused by the last term.
Note that this equation preserves energy conservation in the form $E =
\int u^2 dx = {\rm const}$, but now we have to remember the
transformation leading to (\ref{equ:GOYw}) , i.e. $ u = w k$, where $k
= k_0 (1+\epsilon)^n \approx k_0 e^{\epsilon x}$.  Thus
(\ref{equ:conteps}) can be written
\begin{equation} 
  \frac{\partial }{\partial t}  (w^2 e^{-2\epsilon x}) = 
  -  \frac{\partial }{\partial x}(4 w^3 e^{-2\epsilon x})
  \label{equ:conteps_cons}
\end{equation}
which shows that the appropriate shock-condition is 
\begin{equation} 
V = {\frac{[4 w^3 e^{-2\epsilon x}]}{[ w^2 e^{-2\epsilon x}]}}
 \label{shock}
\end{equation}
where $V$ is the velocity of the shock and $[f(x)]$ denotes the
discontinuity of the quantity $x$ accross the shock. Since $w$ vanish
on one side of the shock we get
\begin{equation} 
V = 4 w
 \label{shock1}
\end{equation}

The characteristic equations for (\ref{equ:conteps}) are
\begin{equation} 
\frac{d x }{d t} = 6 w
 \label{char1}
\end{equation}
and
\begin{equation} 
\frac{d w }{d t} = 4 \epsilon w^2
 \label{char2}
\end{equation}
The latter can be integrated as
\begin{equation} 
  w =  \frac{w_0}{1- 4 \epsilon w_0 t}
  \label{charw}
\end{equation}
where $w_0$ is the initial field taken as a function of the initial
coordinate $x_0$. The second characteristic equation can now be
integrated as
\begin{equation} 
x=x_0 -{\frac{3}{2 \epsilon }} \log (1- 4 \epsilon w_0 t)
 \label{charx}
\end{equation}

Thus each characteristic ends in a singularity at time $t= 1/4
\epsilon w_0(x_0)$. Never the less it is perhaps not obvious that the
shocks also have to end in a singularity at a finite time. To
investigate this, we must investigate the shock condition together
with the solutions (\ref{charw}) and (\ref{charx}). The shock velocity
is $V = dx_e/dt$, where $x_e$ is the edge of the shock. The natural
variable is however $x_0$ and we thus express the shock-condition as
\begin{equation} 
4 w = V = {\frac{dx}{dt}} =  {\frac{\partial x}{\partial x_0}} 
{\frac{dx_0}{dt}}+ {\frac{\partial x}{\partial t}}
 \label{shockx0}
\end{equation}
On performing the differentiations of (\ref{charx}) with $x_0$ and $t$ and 
inserting the solution (\ref{charw}) for $w$ we get
\begin{equation} 
  [1 + (6 {\frac{\partial w}{\partial x_0}} - 4 \epsilon w_0) t ]
  {\frac{dx_0}{dt}}= - 2 w_0
 \label{shockdif}
\end{equation}

To make further progress we must know the form of $w_0$. In our
simulations of bursts, we took distributions like (\ref{Ksol1}), which
are strictly zero outside the interval $[x_1,x_2]$ and approach the
ends of the interval with zero slope. Thus, close to $x_1$ (which must
dominate at large times) the form will be
\begin{equation} 
w_0 \approx a (x_0-x_1)^2
 \label{w0}
\end{equation}
With this assumption (\ref{shockdif}) becomes:
\begin{equation} 
(1 +4 a y (3  -  \epsilon y) t ){\frac{dy}{dt}}= - 2 a y^2
 \label{shockdifa}
\end{equation}
where $y = x_0 - x_1$. When $t \rightarrow \infty$ and $y \rightarrow
0$ the parenthesis on the left hand side must be dominated either by
the term proportional to $t$ or by the constant. If we assume that $ 4
a y (3 - \epsilon y) t \rightarrow 0$ we get the simple equation
\begin{equation} 
{\frac{dy}{dt}}= - 2 a y^2
 \label{shockdifaa}
\end{equation}
with the solution
\begin{equation} 
y = ( c + 2 a t)^{-1}
 \label{shockdifaaa}
\end{equation}
where $c=1/y_0$. With this solution, however, $ 4 a y (3 - \epsilon y)
t \rightarrow 6$ in contradiction to the assumption. Therefore we must
assume that $y t$ does not decay to zero and the dominant terms are
now
\begin{equation} 
3  t {\frac{dy}{dt}}= - y
 \label{shockdifaaaa}
\end{equation}
which gives the decay $y \sim t{-1/3}$. This solution doesn't work
either, since the solution (\ref{charw}) only makes sense
asymptotically if $t y^2 \rightarrow 0$ as $t \rightarrow \infty$. We
therefore conclude that, within the approximation leading to the
continuum limit (\ref{equ:conteps}), the position of the shock will
diverge to infinity at a finite time. It is not clear whether this
behaviour is seen in the discrete equation as well(\ref{equ:weps}).

It is then clear that a pulse reach the dissipative range in finite
time. This behaviour is qualitative the same as was found for the
Parisi equation.

\section{Lyapunov exponents}
\label{sec:lyapunov}
In this section we study the Lyapunov exponents of the GOY model in
the limit of $r \rightarrow 1$. Since we aim at keeping the size of
the inertial range fixed, the number of shells must also be varied as
a function of $r$, if the ``external'' parameters, i.e., viscosity and
forcing are kept fixed at $\nu = 10^{-6}, k_0 = 2^{-4}, f =
(1+i)0.005$. For $r=2$ these parameters correspond to $N=19$ shells;
therefore we use a dependence between $N$ and $r$ as:
\begin{equation}
N = 8+ [ 11{\rm \frac{log(2)}{log(r)}}]
\end{equation}
where the brackets stand for integer value.
We wish to quantify the ``strength'' of the intermittency 
as a function of the effective shell spacing $r$. One way
to do this is by means of the maximal Lyapunov exponent
\cite{kock:98}. 
To compute the maximal Lyapunov exponent in the GOY model, 
we introduce the notation 
${\bf U} \equiv(Re(u_1),Im(u_1),\cdots,Re(u_N), Im(u_N) \, )$
and $F_i = d U_i/ dt$ and
consider the linear variational equations
\begin{equation}
\label{eq2}
\frac{d z_i}{ dt} \ = \sum_{j=1}^{2N}
    A_{ij} \cdot z_j \qquad \ i \ = \ 1,...,2N
\end{equation}
for the time evolution of an infinitesimal increment ${\bf z}=
 {\bf{\delta U}}$,
 where
\begin{equation}
\label{eq3}
A_{nj} \equiv \partial F_n  /  \partial U_j
\end{equation}
is the Jacobian matrix of Eqs. (\ref{un}). The solution for the
tangent vector ${\bf z}$ can thus be formally written as ${\bf
  z}(t_2)={\bf M}(t_1,t_2) \cdot {\bf z}(t_1)$, with ${\bf M}= \exp
\int_{t_1}^{t_2} \ {\bf A}(\tau) d\tau$.  A generic tangent vector
${\bf z}(t)$ is projected by the evolution along the eigenvector ${\bf
  e}^{(1)}$, belonging to the maximum Lyapunov exponent, i.e.  ${\bf
  z}(t) = |{\bf z}(0)| \ {\bf e}^{(1)} \exp(\lambda_1 \ t)$ leading to
\begin{equation}
\label{eq4}
\lambda_1 ~=~ \lim_{t \to \infty} \frac{1}{t} {\rm ln}
\frac{| {\bf z}(t) |}{| {\bf z} (0)|}
\end{equation}
where ${\bf z} (0)$ is the initial tangent vector.

Practically, Eqs. (\ref{un}, \ref{eq2}) are integrated simultaneously
over a certain time $\delta t$, starting with a normalised tangent
vector in a random direction, $\hat{\bf z}(0)$. The increment over
time $\delta t$ in the length of the tangent vector is then $\delta
z_1 = | {\bf z}(\delta t) | / |\hat{\bf z} (0)|$.  Next, the tangent
vector is normalised $ \hat{\bf z} (\delta t) = {\bf z} (\delta t) / |
{\bf z}(\delta t) |$ and this vector is used as a seed for a new
integration over the time $\delta t$ i.e. propagated forward to $t = 2
\delta t$. Generalising this argument we obtain the i'th increment
$\delta z_i = | {\bf z} (i \delta t) | / | \hat{\bf z} ((i-1) \delta
t) |$ and the maximal Lyapunov exponent is given by (where we now set
$\lambda = \lambda_1$):
\begin{equation}
\label{eq5}
\lambda = \lim_{M \to \infty} 
\frac{1}{M}   \sum_{i=1}^M \frac{{\rm ln}(\delta z_i)}{\delta t}
\end{equation}
\begin{figure}[htbp]
  \begin{center}
    \epsfig{file=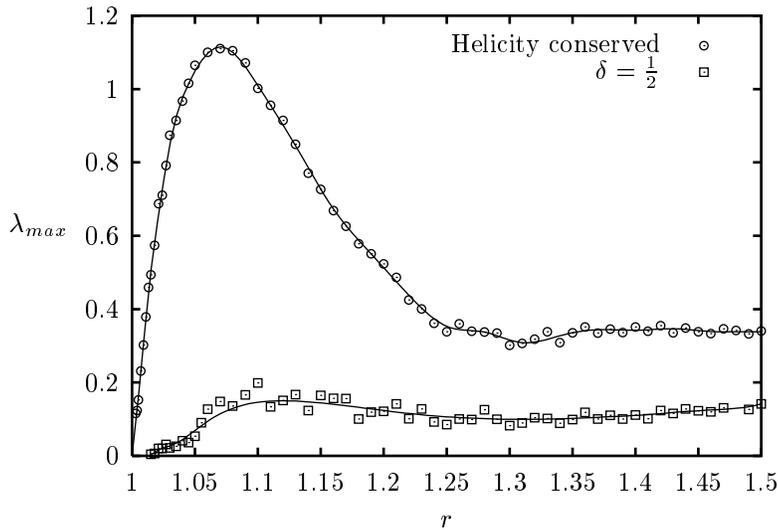} %\input{lyapunov.pstex} 
    \caption{The maximum Lyapunov exponent
      $\lambda_{max}$ for a case with conservation of helicity (i.e.
      $\delta = 1 - 1/r$) and where $\delta$ is fixed at $1/2$. The
      solid line is an interpolation to guide the eye.}
    \label{fig:lyapunov} 
  \end{center} 
\end{figure} 

We have followed two paths for the values of the parameters $a_n, b_n,
c_n$. In the first, the constraint Eq. (\ref{equ:constraint}) is
applied, and in the other we keep the the parameters fixed at the
canonical values $a_n =1, b_n = c_n= - \frac{1}{2}$. The results of
the numerical simulations are shown in Fig. \ref{fig:lyapunov}. In
both cases we observe that the maximal Lyapunov exponent appears to
vanish in the limit $r \to 1$ (the minimal value of $r$ in the plot is
1.004, corresponding to $\sim 1000$ shells).  The lower curve is the
one related to the conservation of both energy and helicity.  The
maximal Lyapunov exponent remains in this case nearly constant in a
large interval of $r$-values and then finally drops to zero. In the
case where only energy is conserved (the upper curves) there is a
maximum around $r=1.1$ after which also this decreases towards zero.
Based on these numerical results we conclude that the GOY model does
not exhibit chaotic dynamics in the zero-spacing limit.  Similar
behavior has been observed recently in a shell model containing two
fields \cite{Biferale}. Here it has been observed \cite{Benzi} that
the intermittency corrections disappear in the limit $r \to 1$.

\section{Conclusions}
The main objective of this paper is to study the dynamical behavior of
single burst motion on a vanishing background in various shell models.
The motion of single bursts is important for the understanding of
intermittent behavior in turbulence. Indeed it is believed that the
presence of intermittency causes the corrections to the classical
Kolmogorov theory which manifest itself as multiscaling of higher
order structure functions.  In the present paper we have studied the
motion of a single burst in three different versions of shell models:

\begin{enumerate}
  \item The standard GOY model when the shell spacing approches, but
    is different, from zero;
  \item The Parisi continuum limit of the GOY model;
  \item The zero-spacing of the GOY model where helicity conservation
    is kept.
\end{enumerate}
In all cases we observe that an initial disturbance splits up in a
left- and a right-moving part. In case 1) the pulse retains its shape and
whereas the right moving part accelerates, the left-moving part
decellerates. In case 2) an initial disturbance splits into left- and
right-moving shocks. The right-moving shock accelerates and its
position diverges to infinity in a finite time. Finally, in case 3)
the shape retains its ``solitary'' form and does not turn into a
shock. Here the velocity also remains constant and the solitons pass
through each other with a slight phase shift but without changing
their shapes, completely as found in integrable field theories.

We have analytically calculated the motion of the shocks in case 2) by
means of characteristics. In case 3) we have analytically estimated
the asymptotic shape of the solitary wave and find that it decreases
super exponentially with a decay constant given by the golden mean. On
a non-vanishing background we have on the other hand found that the
decay of the soliton is a standard exponential. From numerical
simulations in case 1), we have estimated the maximal Lyapunov
exponent and found that it vanishes in the limit of zero-spacing,
although the the motion remains chaotic for finite spacing. This
results is expected since, as we have shown earlier, the continuum
Parisi limit resembles the Burgers equation and gas theory and is in
fact exacly soluble by the method of characteristics.

We hope that our findings could be a starting point for a more
detailed understanding of structure of intermittency, in particular of
how the presence of isolated bursts give rise to corrections to
classical scaling.
 %\section{Conclusions}
%\begin{itemize}
%  \item Mention that we in fact study the case where the middle term
%    vanishes. This term is probably the term responsible for the
%    chaos. 
%  \item For the Helicity-conserving shell model the $r=1$ limit is
%    where the right term vanishes. There the pulses do not travel, but 
%    there are loads of chaos.
%  \item ...
%\end{itemize}

%\ack

%\bibliographystyle{plain}
%\bibliography{shellmodels}

\begin{thebibliography}{1}
\bibitem{Ola:98} We would like to thank Ola Tornkvist and Conor Houghton for
pointing this out to us.
\end{thebibliography}

\end{document}